\documentclass[runningheads]{llncs}
\usepackage[T1]{fontenc}
\usepackage{graphicx}

\begin{document}

\title{High-Resolution Agent-Based Modeling of Campus Population Behaviors for Pandemic Response Planning}
\titlerunning{High-Resolution ABM of Campus Population Behaviors}

\author{Hiroki Sayama\inst{1,2}\orcidID{0000-0002-2670-5864} \and
Shun Cao\inst{3}\orcidID{0000-0001-7014-0728}}

\authorrunning{H. Sayama and S. Cao}

\institute{Binghamton University, State University of New York, Binghamton, NY 13902, USA\\
\email{sayama@binghamton.edu} \and
Waseda University, Tokyo 169-8050, Japan
\and
University of Houston, Sugar Land, TX 77479, USA\\
\email{scao7@central.uh.edu}}

\maketitle
	
\setcounter{footnote}{0}

\begin{abstract}
This paper reports a case study of an application of high-resolution
agent-based modeling and simulation to pandemic response planning on a
university campus. In the summer of 2020, we were tasked with a
COVID-19 pandemic response project to create a detailed behavioral
simulation model of the entire campus population at Binghamton
University. We conceptualized this problem as an agent migration
process on a multilayer transportation network, in which each layer
represented a different transportation mode. As no direct data were
available about people's behaviors on campus, we collected as much
indirect information as possible to inform the agents' behavioral
rules. Each agent was assumed to move along the shortest path between
two locations within each transportation layer and switch layers at a
parking lot or a bus stop, along with several other behavioral
assumptions. Using this model, we conducted simulations of the whole
campus population behaviors on a typical weekday, involving more than
25,000 agents. We measured the frequency of close social contacts at
each spatial location and identified several busy locations and
corridors on campus that needed substantial behavioral
intervention. Moreover, systematic simulations with varying population
density revealed that the effect of population density reduction was
nonlinear, and that reducing the population density to 40-45\% would
be optimal and sufficient to suppress disease spreading on
campus. These results were reported to the university administration
and utilized in the pandemic response planning, which led to
successful outcomes.  
\end{abstract}

\section{Introduction}

The recent COVID-19 pandemic caused an unprecedented level of complex
societal challenges to humanity at all scales, ranging from local
and regional scales to national and global scales. It also created a
significant societal demand for rapid responses from scientific
communities, not only in biomedical and pharmaceutical domains but
also in mathematical, computational and complex systems domains in
terms of modeling, simulation and analysis of epidemic and other
socio-economical dynamics \cite{Siegenfeld2020}. A wide variety of
modeling efforts were made during the pandemic, including traditional
compartmental modeling of the pandemic dynamics
\cite{radulescu2020management,dashtbali2021compartmental}, modeling
and analysis of the pandemic dynamics and its socio-economic impacts
using large-scale human mobility data
\cite{aleta2020modelling,aleta2022quantifying,yabe2023behavioral}, and
agent-based modeling of disease spreading at regional or national
scales (often on synthetic social networks)
\cite{silva2020covid,almagor2020exploring,hinch2021openabm,kerr2021covasim,ventura2022epidemic},
to name a few.

Compared to the popular large-scale disease spreading modeling effort,
there was relatively less effort spent for the modeling of human
population behaviors in small- to mid-sized organization or community
settings. There were some agent-based behavioral modeling studies done
at such small- to mid-scales \cite{wapo,cuevas2020agent}, but those
models were highly styled, simplistic ones without much realistic
physical or social details included. Building a more detailed, more
realistic model of the behaviors of specific small- to mid-scale
populations would be highly useful and informative for critical
decision making of organizational and community administrations (e.g.,
\cite{gressman2020simulating}). In this article, we present a case
study of our own effort to develop a high-resolution agent-based model
of our university campus population. This project was done in response
to the COVID-19 pandemic in May-June 2020\footnote{We were initially
not planning to publish a technical paper on this project because the
model was very specific to the Binghamton University campus and would
not easily generalize to other campuses/communities. However, we have
since kept receiving requests for technical details of this project,
which has convinced us to write this article to share our
experience, almost four years later.}  when our university
administration needed to understand the whole campus-level population
behaviors to help their pandemic response planning and public health
education and outreach. We will describe the background, methodology,
technical details of the developed model, results of systematic
simulations conducted using the model, and how they helped the campus
community decision makers. We hope this report provides useful
documentation that other complex systems researchers may refer to in
case they need to grapple with the next pandemic or other crisis
events in the future.

\section{Background}

At the outbreak of COVID-19 in early 2020, many universities and other
educational institutions faced an unusually difficult problem: to
develop effective mitigation strategies for safely restarting their
educational activities without much knowledge about either
epidemiological properties of the SARS-CoV2 virus or behavioral
dynamics of the individuals on their campuses. Binghamton University
was no exception but, as part of the larger State University of New
York system, it was required by the New York State to safely reopen
the campus by the Fall 2020 semester. The university administration
therefore formed a pandemic response taskforce team, in which one of
the authors (HS) worked on various epidemic modeling/prediction and
data visualization tasks since April 2020 \cite{sayama2021artificial}.

At one of the online taskforce meetings in the early stage of the
pandemic, the university administration realized a need for a detailed
behavioral simulation model of the entire campus population. Using
such a model, they could gain better understanding of people's
behavioral patterns and their interactions so as to make informed
decisions on how much portion of classes should go online, and also to
conduct educational outreach to the campus community on the scientific
rationale for their decisions and the importance of social distancing
and other public health practices. The authors were tasked by the
administration to work on this modeling project on May 15th,
2020. This was an intense, dynamic, time-critical, non-stop modeling
project that lasted for three weeks. All the modeling and systematic
experiments were completed by June 6th, 2020 (total project period: 22
days). This article aims to summarize the experience we gained through
working on this project, which demonstrates how complex systems tools
such as agent-based modeling can be applied to concrete real-world
challenges at local/organizational levels.

\section{Methodology}

It was quite apparent from the beginning of the project that the task
assigned to us was an extremely difficult one, because no direct
empirical data were available for individual students' or employees'
actual behavioral trajectories. Unlike some other regional or national
scale studies, our model was about human movements at much smaller
spatial ($10^0$--$10^3$ meters) and temporal ($10^0$--$10^3$ minutes)
scales. No such high-resolution behavioral data were available for the
target population we were supposed to model, and hence typical data
science/machine learning/AI modeling approaches were not useful at all
in this project.

We therefore tackled this challenge using a mechanistic modeling
approach. Rather than trying to analyze empirical data and find
patterns in them, we aimed at creating a first principle-based
mechanistic model of human individuals' behaviors and thereby
reconstructing a hypothetical (yet reasonably realistic) virtual
campus population in simulation. Each individual agent's behavior was
conceptualized as a migration process on a multilayer transportation
network, in which each layer would represent a different
transportation mode (i.e., on foot or by car/bus).

\begin{figure}[tp]
\centering
\includegraphics[trim={10.3cm 0 8.5cm 0},clip,width=0.82\textwidth]{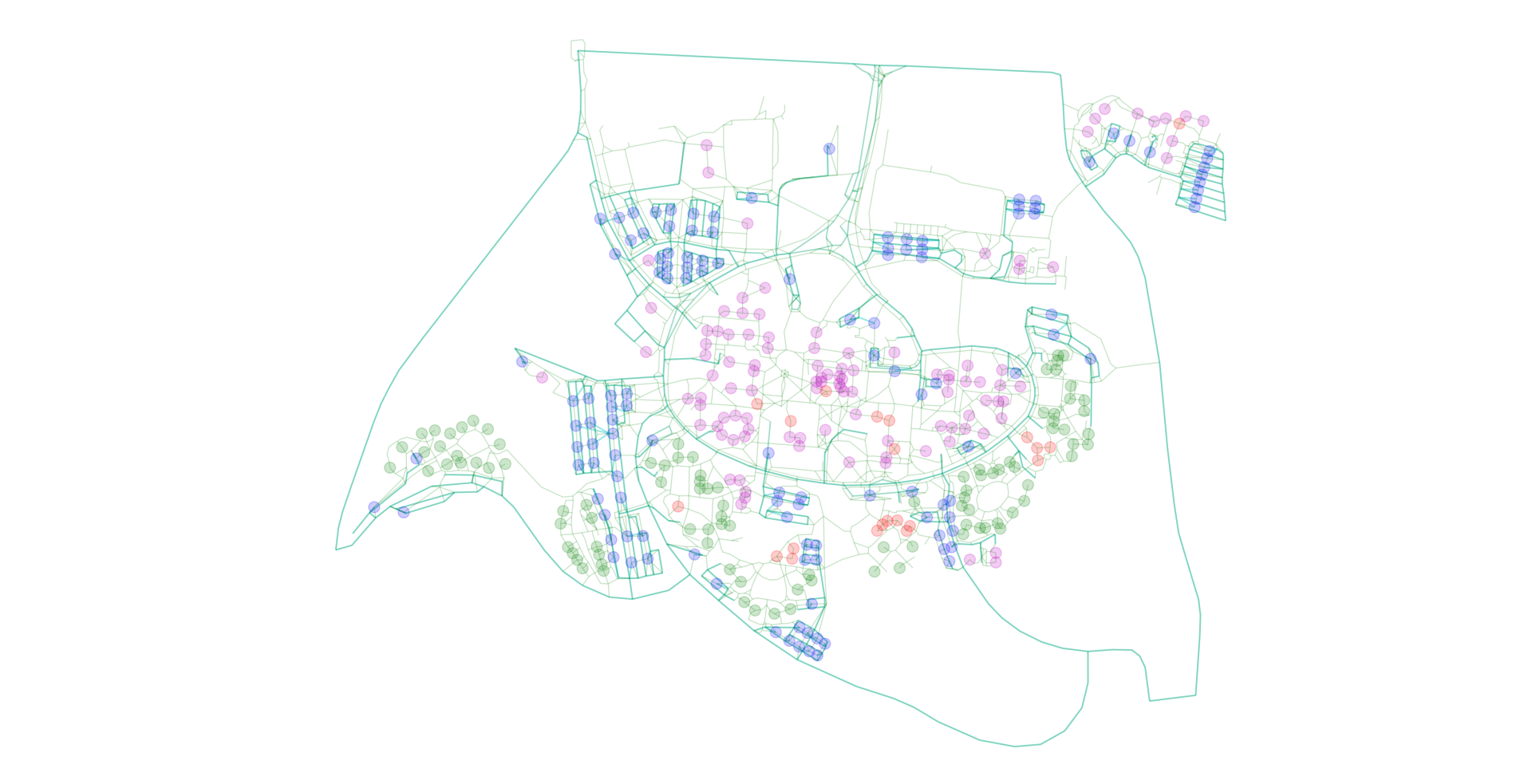}
\caption{The multilayer transportation network of the Binghamton
  University main campus reconstructed using the data manually
  extracted from Google Earth. The first layer of the network (thick lines) are roadways for vehicles,
  whereas the second layer (thin lines) are pedestrian pathways. Pink/orange/green/blue
  circles indicate locations of buildings/food places/residence
  halls/parking spaces, respectively.}
\label{fig:networks}
\end{figure}

To comprehensively map the intricate geographical structure of the
university's main campus, including vehicle roadways, pedestrian
paths, buildings, parking lots, and other facilities, we manually and
meticulously extracted locational data from Google Earth
\cite{google-earth}\footnote{Detailed geographical data were not
directly downloadable from Google Earth.}. We used the ``placemark''
tool \cite{placemark} to collect precise longitude and latitude values
for each point of interest. For both roadways and pedestrian paths, we
recorded adequate spatial data points to accurately depict their
(often curved or bent) trajectories. Similarly, for larger parking
lots, buildings, or facilities with more than one exit, we gathered
multiple location points to provide insight into their internal layout
of pathways. This comprehensive locational information forms the basis
for constructing two distinct networks: a vehicle roadway network and
a pedestrian pathway network. Figure \ref{fig:networks} visualizes
these networks. Within these networks, each node corresponds to a
specific location on campus, while each edge represents the connection
between two adjacent locations on either roadways or pedestrian
pathways. Note that the roadway network is structured as a directed
network, accommodating both one-way and two-way traffic flow
configurations.

To inform the agents' behavioral rules, we collected a wide variety of
indirect information about the campus's daily behaviors from various
divisions/offices of the university, including the following:
\begin{itemize}
  \item Course schedules and allocated classrooms (from Fall 2019)$^*$ 
  \item Locations and sizes/capacities of classrooms (from Fall 2019)$^*$ 
  \item Individual students' residence hall assignments (from Fall 2019, anonymized)$^*$ 
  \item Individual students' course registration records (from Fall 2019, anonymized)$^*$ 
  \item Individual employees' FTEs (full-time equivalents) and office
    locations (from Fall 2019 or Spring 2020?,
    anonymized)$^*$ \footnote{The exact date of this employee dataset
    is uncertain. Also, it was limited only to the university
    employees, because the information about employees of external
    vendors working on campus were not available to the university
    administration.}
  \item Locations of classroom/office buildings and residence halls (from Spring 2020)$^+$ 
  \item Locations of food places (food court, coffee kiosk, etc., from Spring 2020)$^+$ 
  \item Locations of parking lots (from Spring 2020)$^+$ 
  \item Capacities of residence halls (from Spring 2020)$^\dag$
  \item Capacities of parking lots (from Spring 2020)$^\dag$
  \item Hourly occupancy rates of parking lots (from Fall
    2018)$^\dag$\footnote{Only this dataset was from 2018.}
  \item Parking garage and paid lot entries (from Fall 2019)$^\ddag$
  \item Bus routes and arrival/departure times (from Fall 2019)$^\ddag$
  \item Bus boarding/alighting counts by bus stop (from Fall 2019)$^\ddag$
\end{itemize}
The sources of the above data were as follows:
\begin{description}
  \item[$^*$] Office of Course Building and Academic Space Management (CBASM), Binghamton University
  \item[$^+$] Google Earth
  \item[$^\dag$] Division of Student Affairs, Binghamton University
  \item[$^\ddag$] Office of Transportation and Parking Services, Binghamton University
\end{description}

When we were collecting these data, it was not clear to us which data
would be useful or not useful for our modeling purposes; we simply
gathered as much different kinds of data as we could. What was quite
clear, however, was that none of these datasets would provide direct
information about individuals' detailed behaviors and that there would
be no automatic data science/machine learning/AI tools that do magic
to automatically convert them into a reasonable simulation model. We
therefore spent a lot of time and effort to review, clean, and examine
these information pieces to come up with ideas of how we could stitch
them together (as needed and appropriate) to paint a {\em mechanistic}
picture of what would be the most plausible dynamics taking place on
campus. This effort might be somewhat similar to crime scene
reconstruction---no one witnessed the scene directly, but there were
several indirect traces of the crime, and we as the detectives had to
{\em think} a lot ourselves to figure out what would have happened at
the scene. To make this task manageable and practical, we decided to
focus on the campus behavior on a typical Tuesday in a Fall semester
(i.e., not to model for the whole week), and started building a
detailed mechanistic agent-based model, as described below.

\section{Agent-Based Model}

We developed a high-resolution agent-based model that simulates the
detailed movements of individual students/employees on the Binghamton
University campus from early morning until evening on a typical
Tuesday of a Fall semester. The campus was represented as a multilayer
transportation network that consists of the roadway network layer and the pedestrian pathway network layer, as shown in Fig.\ \ref{fig:networks}. The model
simulated a total of about 17,000 individual agents ($\approx$ total
number of individuals that live in or come to the Binghamton
University campus in one day) together with about 10,000 vehicles
($\approx$ total number of individuals that commute from outside the
campus in one day), each of which was simulated with unique behavioral
parameters and attributes estimated and reconstructed according to the
collected data.

We implemented the model in Python 3 using NetworkX
\cite{networkx} for network representation and
visualization. We also used PyCX \cite{pycx} for dynamic interactive
visualization of simulation results. All the coding work was done in a
single plain Jupyter Notebook environment, since the model development
was highly iterative with a lot of experimentations and
parameter/setting adjustments involved and needed to be able to
swiftly respond to continuously changing information and
requirements. The developed model (a Jupyter notebook) and some
animations of simulation results are available from the authors'
GitHub page \cite{github}.

We assumed that each individual agent first develops a plan of the
day, which is a list of ``(time, location)'' pairs that it needs to be
present at according to its class schedule or work duties. This plan
is then expanded to be more detailed about the departure and arrival
times at each location by calculating the estimated transportation
time it needs for each move along the shortest path on a transportation
network. The starting/ending location of the plan of an on-campus
residential student agent is the residence hall they live in, whereas
that of an off-campus student or employee agent is randomly chosen
from the corner intersections of the roadways that surround the
campus, assuming that the agent commutes by car or bus via the chosen
entry/exit point. Those commuting agents plan to get off the bus at
one of the bus stops or park their car at an available parking lot,
where they switch the mode of transportation from driving to walking
(and they do the opposite at the same location when they go back home
later in the day). Those bus stops and parking lots are represented as
common nodes that appear in both layers of the transportation
network. To make the simulation simple, all the commuting agents
pre-select and claim a specific parking space to use beforehand
(i.e., no on-the-spot search for an open parking space).

There are some more individual details in the agents' plans. For example,
on-campus residential student agents may plan to go back to their
residence hall in the middle of the day if there is a sufficiently
long break between classes. All agents plan to go to the nearest food
place at an appropriate lunch time according to their schedule (if
there is enough time to do so). Finally, we also added some individual
behavioral variations in the agents' plans in terms of how fast they
can walk, how punctual they plan to be at the classes, and so on.

\begin{figure}[tp]
\centering
\includegraphics[trim={6.5cm 0 5cm 0},clip,width=0.82\textwidth]{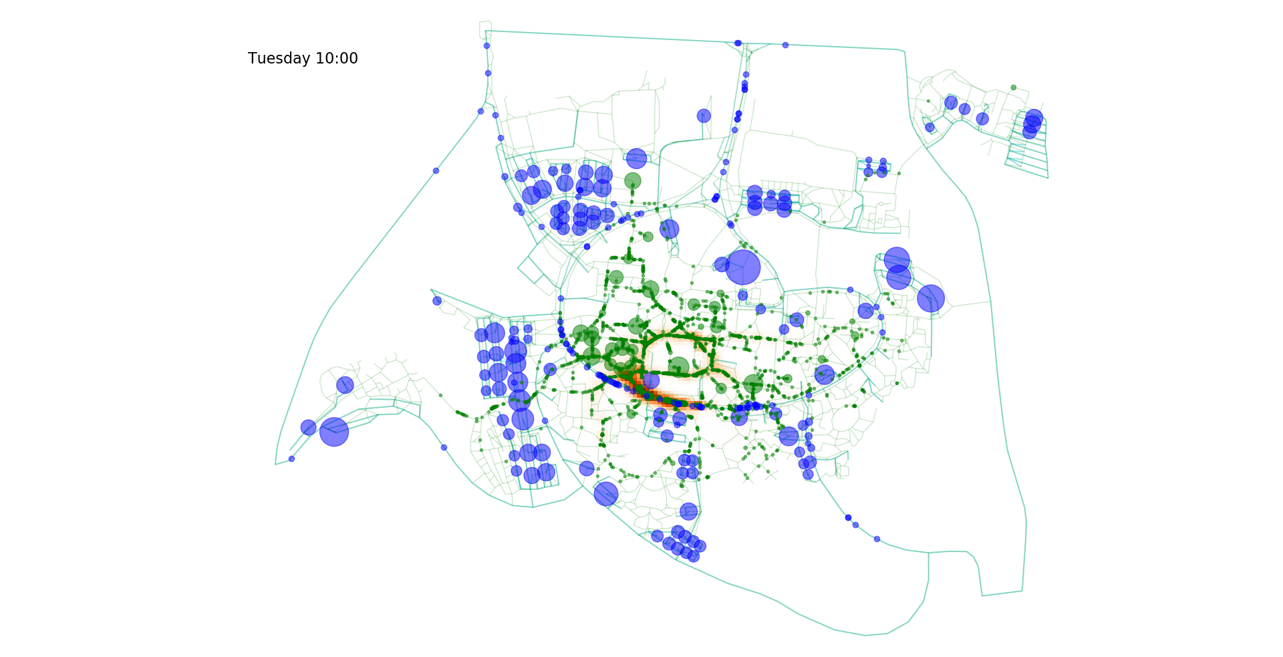}
\caption{A snapshot of a sample simulation run. Tiny green/black dots
  represent student/employee agents, respectively, and small blue dots
  on roadways represent vehicles. Green disks in the central area
  represent buildings, with the disk size scaled according to the
  number of agents inside the building. Blue disks represent parking
  spaces, with the disk size scaled according to the number of cars
  parked there. The red-orange-ish heatmap displayed in the background
  shows the density of close social contacts occurring in each area
  (spatial diffusion and exponential decay operations were applied to
  enhance the visibility and interpretability).}
\label{fig:sample-run}
\end{figure}

Once all the agents complete developing their daily plans, the model
simulates their actual behaviors (entry, movements, exit) in discrete
time steps from 6:30am to 10:30pm (1 step = 1 minute in real life;
total length of one simulation = 16 hours, or 960 steps). An agent's
location is represented by a combination of which edge it is on and
how far it has traversed that edge; this representation allows for
more continuous minute-by-minute movement simulation than simple
edge-hoppings on a network. Furthermore, when an agent is moving on
foot on a pedestrian pathway and too many other agents are nearby, its
speed of movement slows nonlinearly down to 50\%, which simulates
the effect of congestion (which is often seen on campus). This traffic
congestion effect may make agents' behaviors deviate from their
original plans. Figure \ref{fig:sample-run} shows a snapshot of a
sample simulation run, in which the locations of agents (students,
employees, vehicles), building occupancies, and parking occupancies are
visually presented.

This model allowed for detecting and counting the occurrence of close
social contacts, defined as two agents getting closer to each other in
less than six-foot ($\approx$ 2 meters) distance for 10 minutes or
more, as an outcome measure of a simulation. For close social contacts
that would be taking place within an indoor space (a building or a
classroom represented as a single node in the network that has many
agents on top of it), we estimated their number $nCC$ using the
following formula:
\begin{equation}
nCC = \max \left\{\left( \frac{N}{A/a} - 1 \right) \frac{N}{2}, \; 0 \right\}
\end{equation}
Here $N$ is the total number of agents occupying the space, $A$ the
area of the space, and $a$ the area of a three-foot-radius disk (i.e.,
personal space that should be maintained clear). The above formula
roughly estimates the average number of close neighbors each agent has
($N/(A/a) - 1$) and then multiplies it by the number of agents, divided
by 2 (since each close contact would be counted twice in two
directions). These detected or estimated close social contacts were
counted cumulatively in a $200 \times 200$ spatial cellular grid and
visualized as a background heatmap as well
(Fig.\ \ref{fig:sample-run}). Spatial diffusion and exponential decay
operations were also implemented on this background spatial grid, just
to enhance the visibility and interpretability of the spatio-temporal
dynamics of close social contacts.

\begin{figure}[tp]
\centering
\begin{tabular}{l}
(a)\\
  \includegraphics[trim={6cm 0 6cm 0},clip,width=\textwidth]{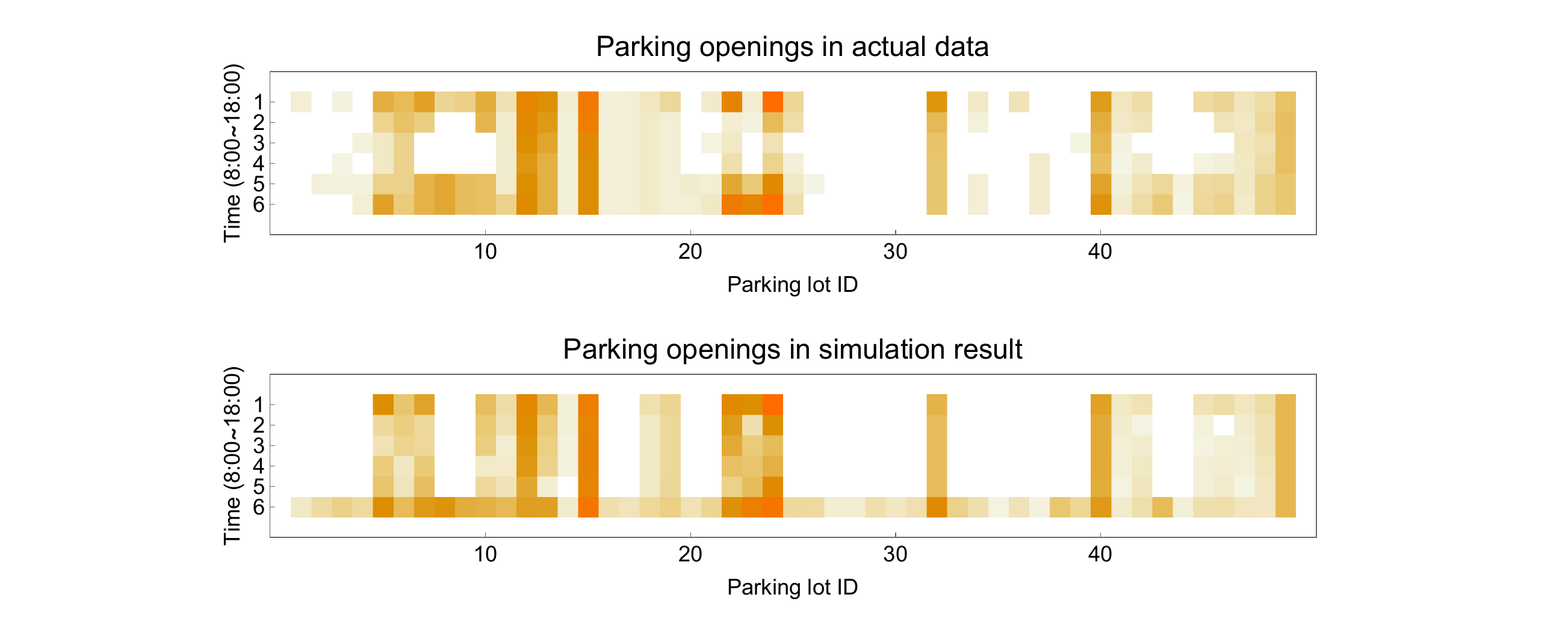}
\end{tabular}
~\\
\begin{tabular}{l}
(b)\\
\includegraphics[width=0.7\textwidth]{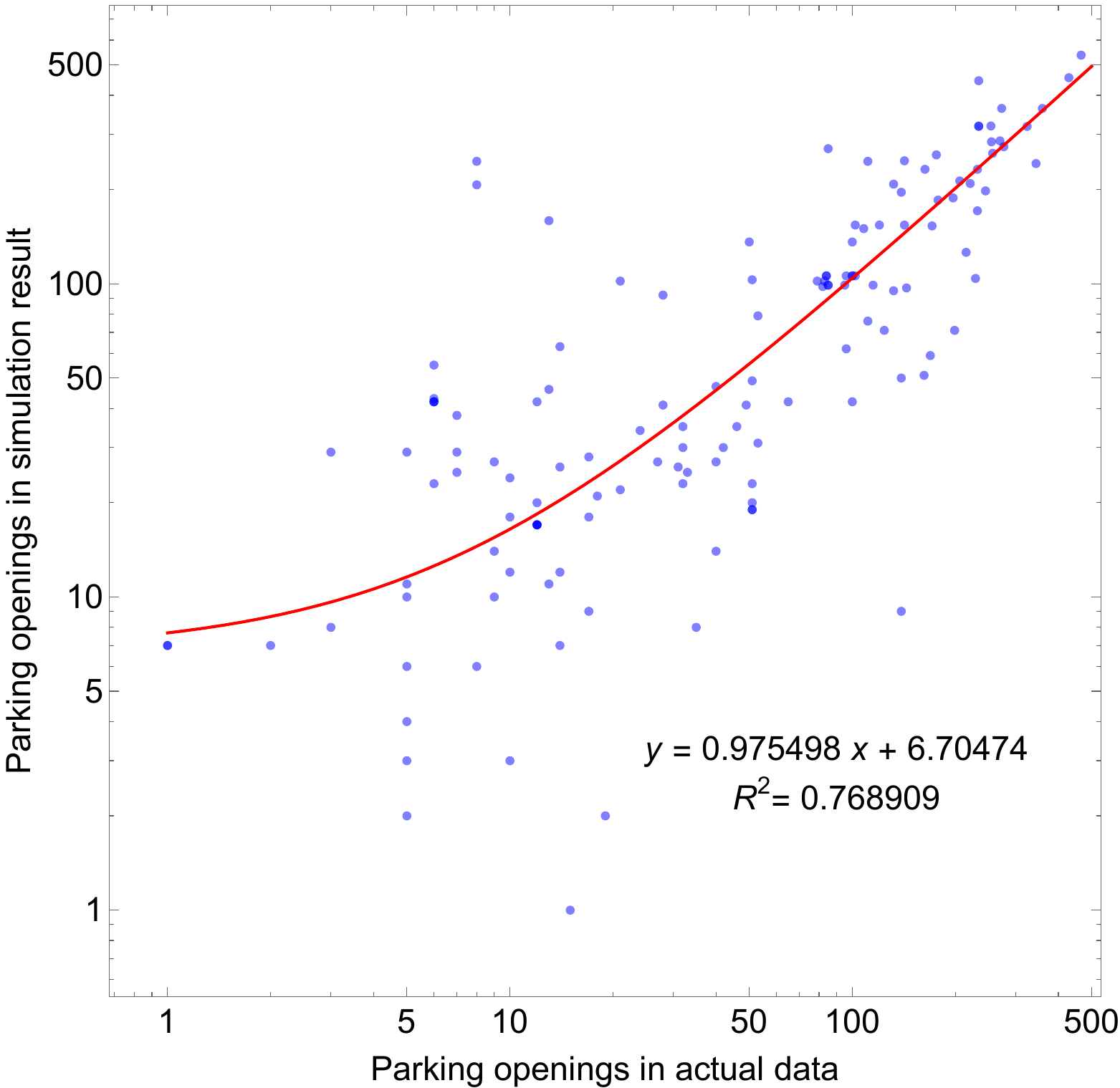}
\end{tabular}
\caption{Comparison of the numbers of parking openings between the
  actual data and the simulation result using a calibrated
  model. (a) Heatmaps showing the number of parking openings along
  space (x-axis: parking lot ID) and time (y-axis: two-hour time
  points from 8:00am to 6:00pm). The top and bottom maps visualize the
  actual data and the simulated result, respectively, which show
  reasonable similarity. (b) Result of linear regression between
  the actual data and the simulated result (plotted in log-log scales,
  which makes the fitted linear line curvy). Each blue dot represents
  values for a specific space-time pair. The linear line fit shows
  high correlation ($R^2 = 0.768909$).}
\label{fig:parking}
\end{figure}

Before conducting a set of systematic numerical simulations, we
manually explored and calibrated various model assumptions and
parameters so that the model behavior would match the reality
reasonably well. For this purpose, we chose the daily parking
occupancy rate as the target variable to use to evaluate the model
validity, as it was the only observational behavioral data that we could
compare the model with. Figure \ref{fig:parking} shows visual
comparison (Fig.\ \ref{fig:parking}(a)) and statistical regression
(Fig.\ \ref{fig:parking}(b)) between the actual data and the simulated
result of the parking openings in each parking lot and every two-hour
time point from 8:00am to 6:00pm. The simulated result showed a
reasonably high correlation with the actual data ($R^2 = 0.768909$),
which gave us some assurance that the developed model would not be so
far off from the reality.

More details of the model can be found in the simulator code posted to
the author's GitHub page \cite{github}.

\section{Results}

\begin{figure}[tp]
\centering
\includegraphics[trim={6cm 0 4cm 0},clip,width=\textwidth]{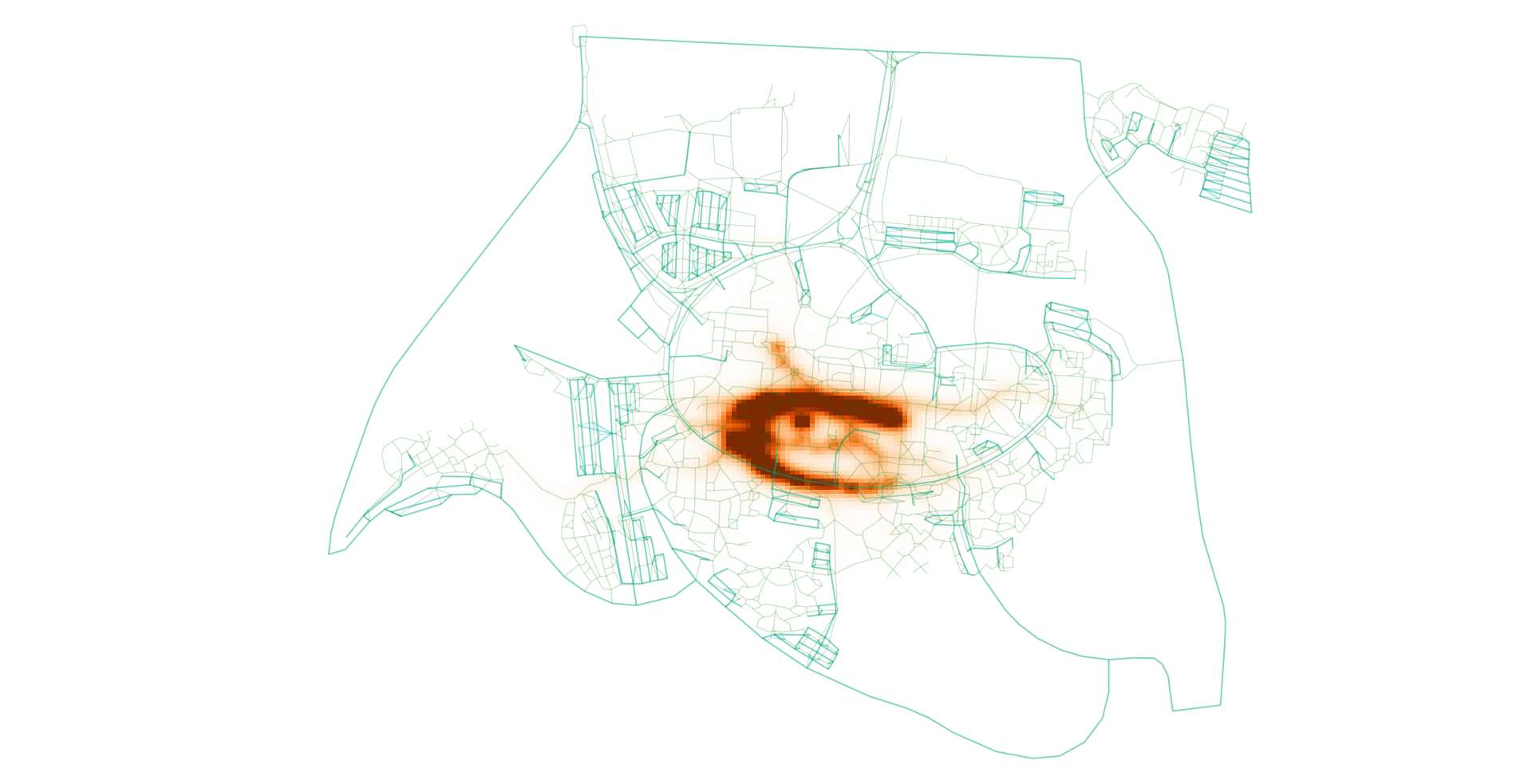}
\caption{Summary density map visualizing the cumulative frequencies of
  close social contacts for the whole campus population over the
  entire duration of simulation (a whole day). A ``C''-shaped
  high-contact area is clearly visible, which connects the food court
  (top right of the ``C'') to the Lecture Hall (left of the ``C'') and
  then to the bus terminal (bottom right of the ``C''). Another
  clearly visible dot inside the ``C'' corresponds to the Library
  Tower, which has a popular coffee kiosk on the ground floor.}
\label{fig:density-map}
\end{figure}

Using the calibrated model, we conducted a series of simulations of
the campus population behaviors. Figure \ref{fig:density-map} shows a
summary density map that visualizes the cumulative frequencies of
close social contacts for the whole campus population over the entire
duration of simulation (a whole day). This map clearly shows several
busy locations and corridors in the campus that would experience high
frequencies of close social contacts, where appropriate behavioral
intervention would be needed.

\begin{figure}[tp]
\centering
\begin{tabular}{cc}
\includegraphics[trim={10.3cm 0 8.5cm 0},clip,width=0.49\textwidth]{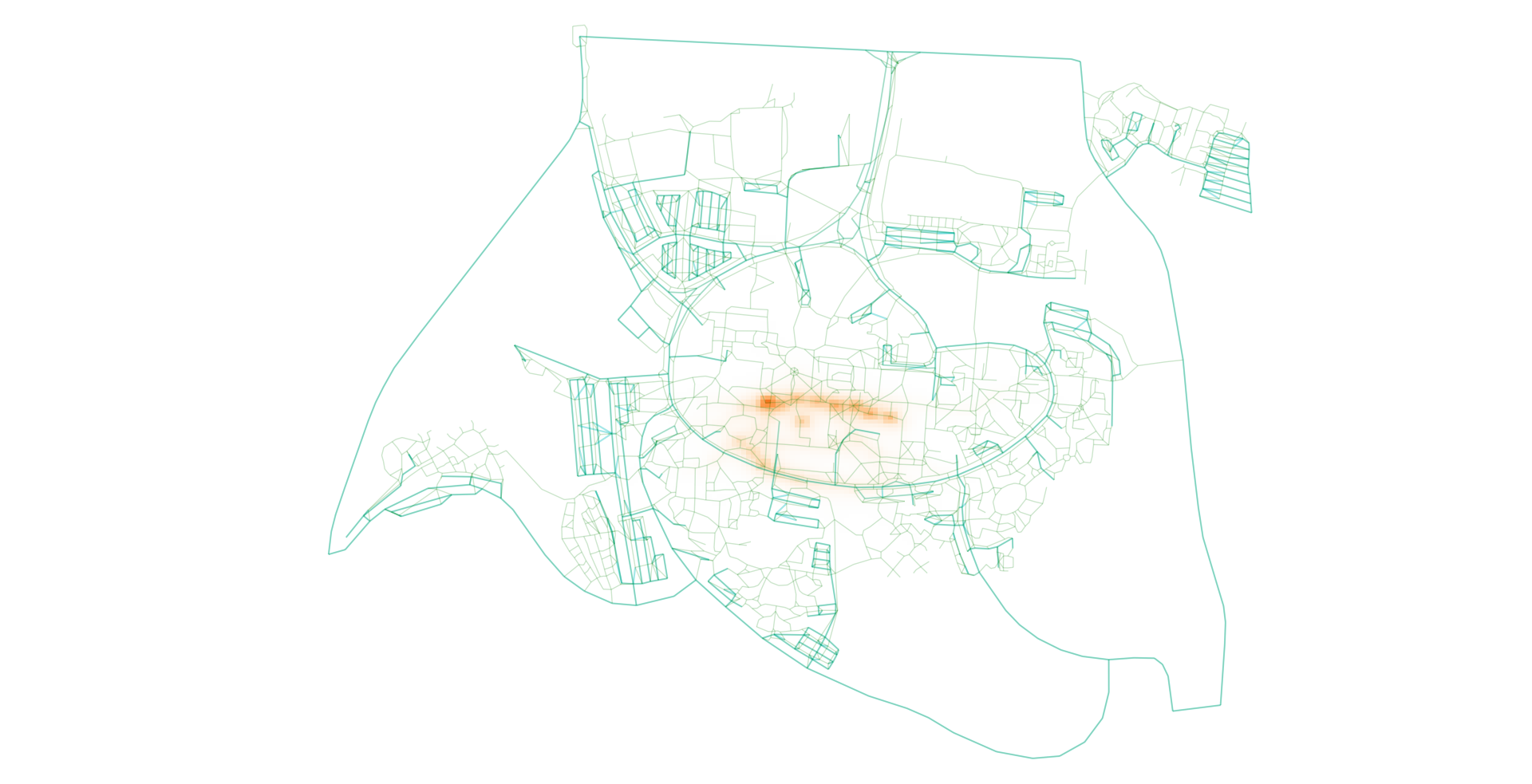} &
\includegraphics[trim={10.3cm 0 8.5cm 0},clip,width=0.49\textwidth]{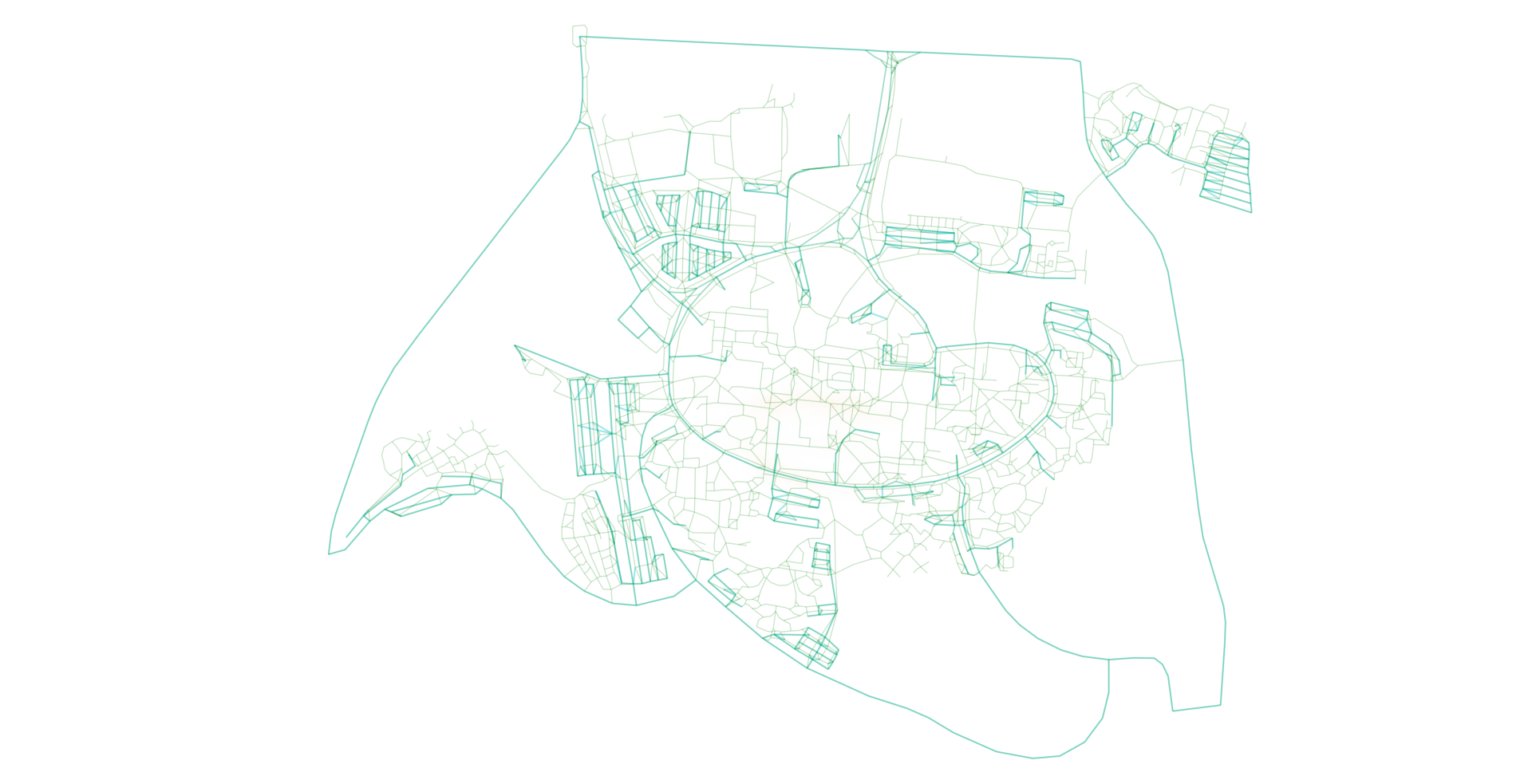}\\
(a) 50\% population density & (b) 25\% population density
\end{tabular}
\caption{Summary density maps visualizing the cumulative frequencies
  of close social contacts with reduced population density. (a) 50\%
  of normal population density. (b) 25\% of normal population
  density.}
\label{fig:density-map-50-25}
\end{figure}

Moreover, having a mechanistic agent-based model enabled us to conduct
systematic simulations with varying population density. Figure
\ref{fig:density-map-50-25} shows summary density maps obtained with
reduced population density by randomly removing a certain fraction of
agents from the simulation (Fig.\ \ref{fig:density-map-50-25}(a): 50\%
population density, Fig.\ \ref{fig:density-map-50-25}(b): 25\%
population density). These results show that reducing the number of
individuals present on campus clearly helps reduce the risk of
disease spreading via close social contacts.

\begin{figure}[t]
\centering
\includegraphics[width=\textwidth]{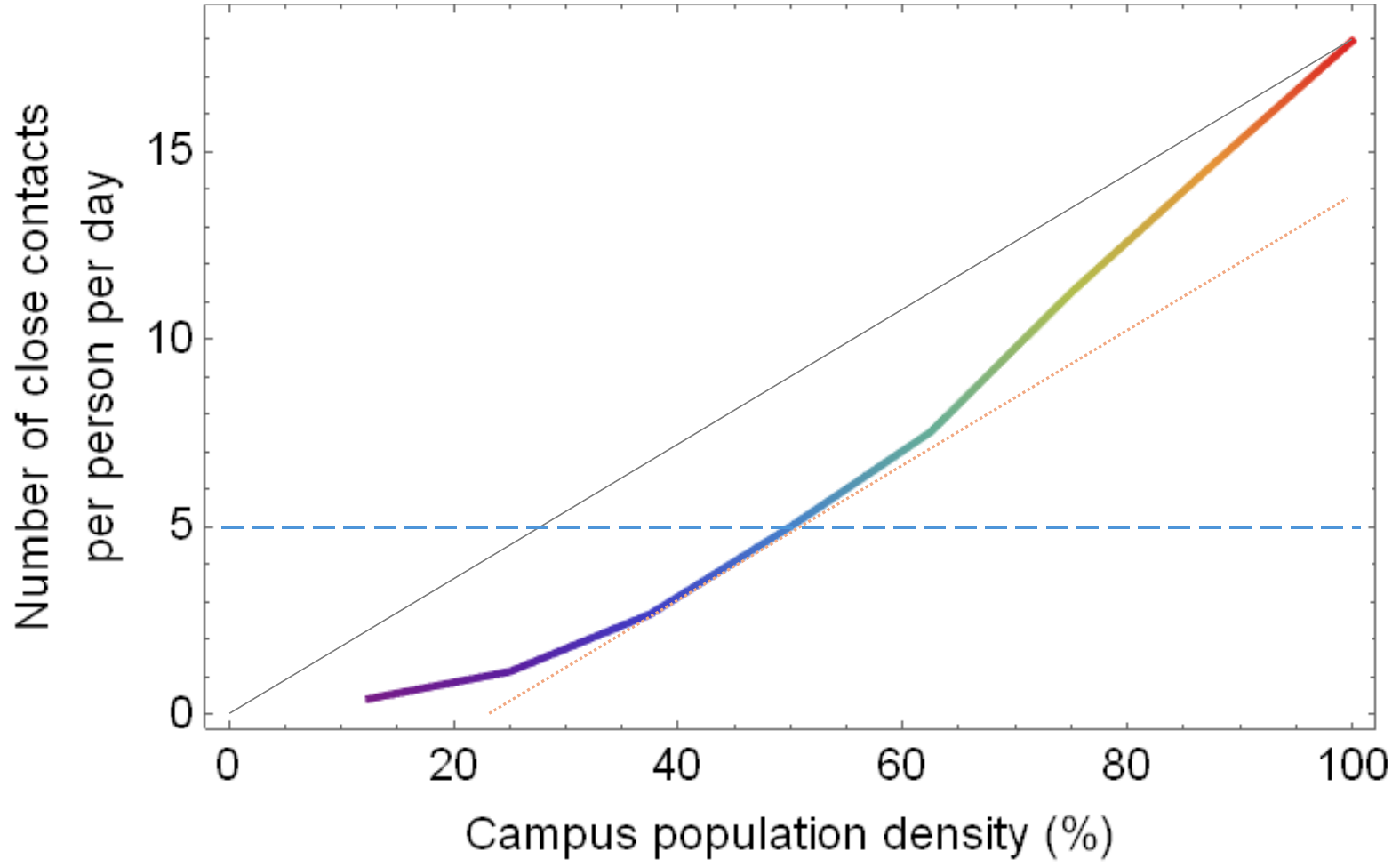}
\caption{Results of population density sweep simulations. The thick
  curve shows the result of the simulations (color gradation added to
  visualize the risk of disease spreading). The grey diagonal line is
  a reference showing a hypothetical linear scenario in which $x$\% of
  population density reduction results in $x$\% of reduction of close
  social contacts. The orange dotted line is parallel to the grey
  reference line, showing the economically optimal point (about
  40-50\%). The blue dashed horizontal line shows an epidemiologically
  estimated threshold value of close social contact frequency at which
  the basic reproduction number of the disease is 1. See text for more
  details.}
\label{fig:density-sweep}
\end{figure}

Figure \ref{fig:density-sweep} summarizes the results of population
density sweep simulations, plotting the average number of close social
contacts per individual agent per day as a function of the campus
population density. This result revealed that the effect of population
density reduction was nonlinear. From an economical viewpoint
considering the return on investment (Fig.\ \ref{fig:density-sweep},
orange dotted line), reducing the population density down to 40-50\%
of the normal level would produce more benefits (i.e., percentage
reduction of close social contact frequency) than investments made
(i.e., percentage reduction of population density). We also considered
the critical value of close social contacts per individual per day
from an epidemiological viewpoint. More specifically, we used the
then-available conservative estimates of the duration in which a
COVID-19 patient remains asymptomatic yet infective ($d = 4$ days;
source: \cite{kucirka2020variation}) and the COVID-19 infection probability
per close social contact ($p_i = 0.05$; source:
\cite{jing2020household}). With these, the basic reproduction number
$R_0$ of the disease can be represented as
\begin{equation}
R_0 = d n p_i ,
\end{equation}
where $n$ is the number of close social contacts per individual per
day. Letting $R_0 = 1$ gives $n = 5$ (Fig.\ \ref{fig:density-sweep},
blue dashed line), implying that if each agent experienced fewer than
five close social contacts per day, then the disease would likely not
spread within the campus.

Based on these results, we concluded that reducing the campus
population density to 40-45\% of the normal level would be optimal and
sufficient to suppress disease spreading within the campus. These
outcomes as well as several visualization and presentation materials
were submitted to the university administration. By considering them
as well as a number of other logistic, administrative and educational
factors, the administration eventually developed a plan to make a
number of courses go online and operate the campus at about 40\% of
the normal population density. Also, the presentation produced from this
project was used for communication and public outreach purposes
\cite{newsarticle,youtubevideo} to provide the scientific rationale
for the administrative decisions and also to increase the awareness of
the importance of social distancing and other public health
practices. As a result, thanks to this and countless other COVID-19
mitigation efforts implemented on campus, Binghamton University
successfully completed the Fall 2020 semester without a major cluster
infection incident or a full campus shutdown.

\section{Conclusions}

In this article, we documented a case study of an application of
high-resolution agent-based modeling to the pandemic response planning
of a university at an early stage of the COVID-19
pandemic. This illustrates how complex systems researchers can make
significant contributions to societal/organizational decision making
and policymaking at a crisis time with great uncertainty. In
particular, we emphasize the importance of creating a causal,
mechanistic model of complex systems based on logic, common-sense
reasoning and first principles. Such an approach becomes extremely
powerful and necessary especially when society faces an unprecedented
situation for which little to no data or prior knowledge would be
available (and thus data science/machine learning/AI methodologies
would not be helpful). In such situations, gaining qualitative insight
and exploring various hypothetical scenarios using mechanistic models
can be much more valuable and more practical than trying to make
accurate, well-validated predictions using a very limited amount of
data. The crisis will not wait until researchers gain an enough amount
of data to conduct rigorous quantitative studies.

It should be pointed out that the model developed in this
project was far from perfect and it ignored a large number of
important factors, including (but not limited to):
\begin{enumerate}
\item Physical details and interactions among agents
\item Width/multiple lanes of pathways
\item Size of roadway/pathway intersections
\item Detailed locations of classrooms, offices and dormitory rooms inside each building
\item Adaptive re-routing behaviors of agents when they face congestion
\item Close social contacts inside dormitory rooms
\item Morning/evening behaviors (e.g., having breakfast/dinner at
  dining halls, after-five partying)
\item Detailed bus schedules and routes
\item Commuters' behavior to adaptively search for an open parking space
  on the spot
\item Traffic to/from the outside of the campus during the day
\item Other campus locations
\end{enumerate}
We also note that our model was highly specific to the Binghamton
University campus and its unique spatio-temporal structures, which
would not be easily generalizable to other university campuses or
facilities. It would be rather difficult to develop a general-purpose
simulation platform for similar modeling projects because the relevant
variables, factors, and available data will greatly depend on specific
local situations, and the information needed by local decision makers
will also vary greatly. We believe what is most important is to
develop a highly collaborative, adaptive, agile multidisciplinary team
in which complex systems modelers work closely with administrators,
decision makers, public health professionals, social science
researchers, and other experts. We ourselves gained a lot of valuable
experience through such agile multidisciplinary collaboration in this
project. We hope the general approach and take-home lessons summarized
in this article can be helpful for other researchers and practitioners
to be prepared for future pandemic or other crises.

\begin{credits}

\subsubsection{\ackname}
The authors thank the Binghamton University administration for
financial support and provision of various data that were utilized in
the modeling.

\subsubsection{\discintname}
The authors have no competing interests to declare that are relevant
to the content of this article. The content and opinions expressed in
this article are strictly of the authors only and have not been reviewed or
approved by Binghamton University.

\end{credits}

\bibliographystyle{splncs04}
\bibliography{sayama-ccs-ppam2024}

\end{document}